\documentclass{aa}
\usepackage{epsfig}
\usepackage{graphicx}
\begin{document}
 
\title{Warps and correlations with intrinsic parameters of galaxies in the visible and radio}
\author{N. Castro-Rodr\'\i guez\inst{1,2} \and M. L\'opez-Corredoira\inst{2}
\and M.L. S\'anchez-Saavedra\inst{3} \and E. Battaner\inst{3}}
\institute{Instituto de Astrof\'{\i}sica de Canarias, E-38205 La Laguna, Spain
\and
Astronomisches Institut der Universit\"at Basel, Venusstrasse 7, Binningen, Switzerland
\and
Departamento de F\'\i sica Te\'orica y del Cosmos, University of Granada, Avd. Fuentenueva SN., E-18002,
Granada, Spain}

\offprints{ncastro@astro.unibas.ch}

\date{vers. May 27 2002 / Received xxxx / Accepted xxxx}

\abstract{From a comparison of the different parameters of warped galaxies
in the radio, and especially in the visible, we find that:\\
a) No large galaxy (large mass or radius) has
been found to have high amplitude in the warp,
and there is no correlation of size/mass 
with the degree of asymmetry of the warp.\\
b) The disc density and the ratio of dark to luminous mass
show an opposing trend: smaller values give more asymmetric warps
in the inner radii (optical warps) but show no correlation with the 
amplitude of the warp; however, in the external radii neither is there any
correlation with the asymmetry. \\
c) A third anticorrelation arises from a comparison of
the amplitude and degree of asymmetry in the warped galaxies.\\
Hence, it seems that very massive dark matter haloes
have nothing to do with the formation of warps but only
with the degree of symmetry in the inner
radii, and are unrelated to the warp shape for the outermost radii.
Denser discs show up the same dependence.
}

\titlerunning{Warps in optical and radio data}

\maketitle

\section{Introduction}

Many spiral galaxies have warps, disc distortions 
with an integral-sign shape (S-warp), cup-shape (U-warp), or
some form of asymmetry.
The Milky Way is an example (Burton 1988; 1992). Indeed,
most of the spiral galaxies for which we have relevant information
on their structure (because they are edge on and nearby) 
show a warp. S\'anchez-Saavedra et al. (1990, 2002) and 
Reshetnikov \& Combes (1998) show that nearly half of the 
spiral galaxies of selected samples are
warped, and many of the rest might also be warped since warps 
in galaxies with low inclination are difficult to detect.
For high redshift, it seems that the effect of warping is even
stronger (Reshetnikov et al. 2002).

At present, there are several theories in the literature about the
causes of warps in galactic discs. 
Four remarkable examples of theories which explain the formation of warps
are:

\begin{itemize}

\item Gravitational tidal effects on a given spiral galaxy due to the 
presence of a satellite. This does not seem to be enough to induce the observed 
amplitude in the Galactic warp (Hunter \& Toomre 1969).
Weinberg (1998) proposed a mechanism for amplifying the tidal effects
caused by a satellite by means of an intermediate massive halo around
the galactic disc, but Garc\'\i a-Ruiz et al. (2000)
have found that the orientation of the warp is not compatible with
the generation of warps by means of this mechanism if the satellites
are the Magellanic Clouds, and moreover the magnification of
the amplitude is not so high (Garc\'\i a Ruiz 2001, ch. 2).
However, this mechanism could operate in galaxies other than the
Milky Way.

\item The intergalactic magnetic field has been suggested as the cause of
galactic warps (Battaner et al. 1990;
Battaner et al. 1991; Battaner \& Jim\'enez-Vicente 1998) directly affecting
the gas in the Galactic disc and producing warps in it.
Stellar warps in the old population could also be possible as a result of
interactions between the gaseous and stellar discs.
This gives rise to some interesting predictions, such as the 
alignment of warps of different galaxies (Battaner et al. 1991) and 
differences between the stellar and gaseous warps.

\item 
Cosmic infall is invoked to explain the reorientation of
a massive Galactic halo, which produces a warp
in the disc (Ostriker \& Binney 1989; Jiang \& Binney 1999). 
This model requires a halo that is much more massive than the
disc, an extremely high accretion rate
(3 disc masses in 0.9 Gyr; Jiang \& Binney 1999) and, 
in this scenario, after a sufficiently long
time, the angular momentum of the Galaxy to become parallel
to the direction of the falling matter causing the warp to decay.
It is difficult to understand why the warps are so frequent in this scenario
but this difficulty might be overcome by including a prolate halo 
(Ideta et al. 2000), which would prolong the warp's existence. 
Also, it is difficult to understand how a low-density halo can retain the accreted intergalactic matter.
A generic misalignment between halo and the disc 
(Debattista \& Sellwood 1999) might answer the question, but
then we need to think about the reason for the misalignment.

\item The accretion of the intergalactic
medium directly on to the disc is another possibility (Revaz \& Pfenninger 2001; L\'opez-Corredoira
et al. 2002). The torque produced in the different rings of the disc
by an intergalactic flow of velocity $\sim 100$ km s$^{-1}$ and baryon density 
$\sim 10^{-25}$ kg m$^{-3}$ is enough to generate the observed warps (L\'opez-Corredoira
et al. 2002) and also predicts the existence of U-warps (cup-shaped), 
which are 
less frequent than S- (integral-shaped) warps. Some alignment of the warps
in neighbouring galaxies and differences between the gaseous and stellar warps 
might also be expected. This hypothesis assumes only
the existence of a not necessarily homogeneous intergalactic medium
with a reasonably low density. Although the idea is plausible and
compatible with our observations, there are still no direct proofs of the
existence of this intergalactic medium.

\end{itemize}

All these theories make different
assumptions about the conditions of the spiral galaxies and their 
neighbourhood (massive haloes, magnetic fields, intergalactic medium, satellites), 
so the study of warps becomes interesting as a tool for discriminating 
among the different scenarios. There are already important works
about the observed properties of warps, however we feel that
these are inadequate, and that an effort must be made to
reduce the number of possible hypotheses. Here, we do not aim to give a final answer
to the question but instead present some new correlations
that might be useful together with other data to discriminate 
among the different models.

Some interesting observational results already published are
the dependence on the environment (isolated or in clusters) 
of the warp amplitude asymmetries (differences between the east and west
wings of the warp) and frequency (in the visible [Reshetnikov \& Combes 1998] 
and also in the radio [Garc\'\i a-Ruiz 2001; Kuijken \& Garc\'\i a-Ruiz 
2001]). Curiously, more isolated galaxies seem to
be more frequently warped (Garc\'\i a-Ruiz 2001; Kuijken \& Garc\'\i a-Ruiz
2001) and this would be
against the gravitational interaction of satellites, at least in some cases. 
Halo-disk misalignments without external dependence are also excluded. 
It seems that intergalactic magnetic fields, or the accretion
of intergalactic matter on to either the halo or the disc are better
representations.   
Therefore, there already exists in the literature some papers which have correlated the warp characterictics with
the environmental parameters. We are not going to explore these correlations
again, but the correlations with the intrinsic
parameters of galaxies. For instance, one interesting question now is whether there is any correlation
between halo properties and warp amplitude/asymmetry. 

If the halo were an important element in the formation of warps,
we should observe some dependence on it. Some models have a warp amplitude depending
on the halo mass. Apart from the hypotheses which talk about a halo as an intermediary between
external forces and the disc (Weinberg 1998: Jiang \& Binney 1999),
works by Nelson \& Tremaine (1995) or Debattista \& Sellwood (1999) predict that, although in most cases the dynamical
friction between the disc and the halo damps the warp, it can also excite the warp.
This is the reason, why we shall try to analyze the optical and radio warps 
in the present paper through the correlations
with the mass/luminosity ratios derived from the rotation curves
(provided they are related to the fraction of dark matter in the galaxies). 
We also produce correlations with other
parameters that represent the intrinsic size of the galaxy
(radius, mass, or luminosity).

\section{Data}
\label{.data}

This study is based on two samples of galaxies, one of 228 galaxies in optical 
bands (see Table 1) and the other of 26 galaxies in radio. We have completed the information
on the warp amplitudes with some intrinsic parameters of the galaxies.

\subsection{Optical data}

The optical warp measurements come from
S\'anchez-Saavedra et al. (1990, 2002), who have analysed images obtained 
from the Palomar Observatory Sky Survey (POSS) and the DSS. 
They measured the amplitude of the warp in galaxies
mostly from the southern 
hemisphere (S\'anchez-Saavedra et al. 1990, 2002);
however, we also took some data from the northern hemisphere
(S\'anchez-Saavedra et al. 1990). 
The galaxies were selected according to the following criteria:

\begin{itemize}

\item High surface brightness ($B<14.5$).
\item Galaxies large enough to detect the warp. The galaxies have a
log$R_{25}$ $\ge$ 24 arcsec ($R_{25}$ is the radius of the angular 
size of the isophote $\mu$ = 25 mag/arcsec$^{2}$).
\item Morphological type, $T$, between 0 and 7.
\item Inclination angle greater than 75 degrees.

\end{itemize}

We took these data and sought several intrinsic parameters of each
galaxy in the literature. 
The $H$-band ($\lambda$ 1.6 $\mu$m) magnitude was used because 
it is a good mass tracer of the stellar population. 
Moreover, the luminous mass in NIR bands 
is not much affected by dust and gas extinction as the visible bands are
and is less contaminated by the young population of the spiral arms.
The total luminosity of a galaxy near 2 $\mu$m is thought to be a
better tracer of the stellar mass than the visible (which is biased by recent
star formation) or the far-infrared (biased again by recent
star formation, which creates and heats dust grains that emit thermally
in this waveband; Jablonka \& Arimoto 1992).
We then tried to
find some correlations between these intrinsic parameters and the warp amplitudes 
in our sample. In Table 1 are shown all the galaxies with these parameters. 
The columns list the following information:

\begin{itemize}

\item Column 1: PGC and NGC number.  

\item Columns 2 and 3: Warp amplitudes extracted from S\'anchez-Saavedra et al.
(1990, 2002). The first number is the amplitude on the east 
side of the galaxy and the second number is the warp amplitude on the west side. 
These values divided by 100 give the tangent of the angle. This is 
the angle between the galactic centre and the end of the warp.

\item Column 4: Redshift of each galaxy taken from the NED. Each reference is 
specified in the table.

\item Column 5: Rotation velocity of the galaxy at $R_{25}$, which
is more or less the maximum rotation velocity. This parameter
comes from several sources, mainly from Mathewson et al. (1992)
and Persic \& Salucci (1995). Each reference is specified in the columns of the
table. 

\item Column 6: the values of $\log D_{25}$ extracted from the LEDA database, 
where $D_{25}$ is the diameter of the isophote with 25 mag/arcsec$^{2}$ in units 
of 0.1 arcmin.

\item Column 7: Magnitude in $H$ from 2MASS (Jarret et al. 2000) and
de Vaucouleurs \& Longo. (1988).

\item Column 8: Distance to the galaxy, obtained using the redshift
value and the Tully--Fisher relation with a Hubble constant of 75 km s$^{-1}$ Mpc$^{-1}$. This method is not
very accurate for very close galaxies; therefore, when the redshift
is less than 1000 km s$^{-1}$, we used distances from Huchtmeier \& Richter (1989).

\end{itemize} 

With this information, we sought any correlation between the amplitude 
of the warp or the difference 
between the east side and west side, and mass/luminosity, dimensions,
infrared luminosity or total mass of the galaxies derived from the rotation
curves. 
The results are commented on in \S \ref{.results}.

\subsection{Radio data}

Radio data are from Garc\'\i a-Ruiz (2001). There are only 26 galaxies
with measurements of warps at these wavelengths.
All the information is given in Table 2; the meaning of each column
is the same as for the optical data. 
There are no other important works on radio warp
amplitudes in the literature. In most cases, the warp is more prominent 
in radio observations than
in optical images because the former extends to greater galactocentric distances. 
The galaxies were selected according the following criteria:

\begin{itemize}

\item Listed in the {\itshape Upsala General Catalogue of Galaxies} (Nilson 1973).
\item Galaxies from the norther hemisphere with declinations higher than 20
degrees.
\item Blue diameters greater than 1.5$'$.
\item Optic inclination angles greater than 75 degrees.
\item Flux density higher than 100 mJy in the radio.
\end{itemize}

Again, we have performed the same analysis as in the previous section for optical 
data, as described in the following section.

\begin{figure*}
\begin{center}
\mbox{\epsfig{file=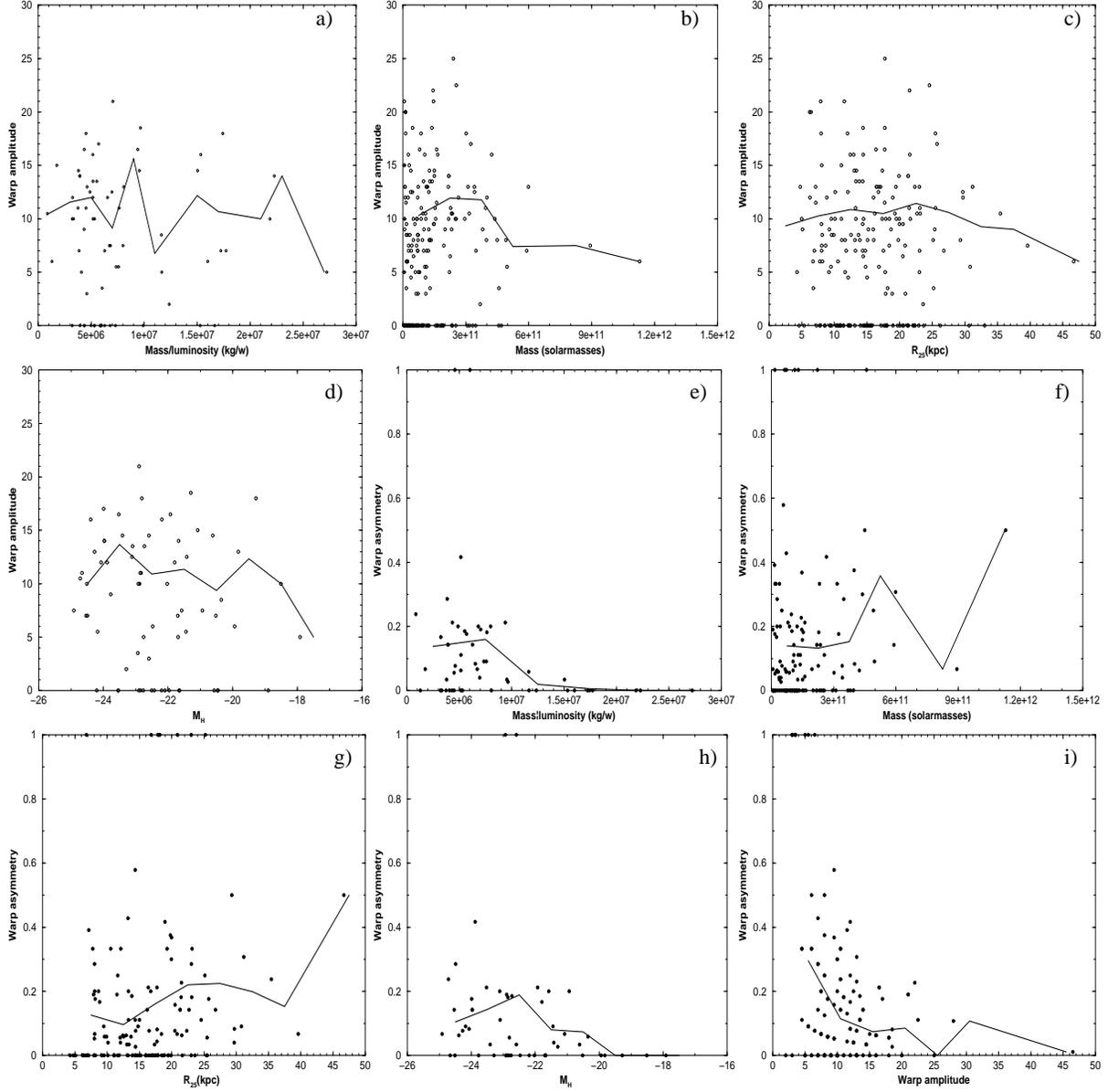,width=16.cm}}
\end{center}
\caption{Optical data from S\'anchez-Saavedra et al. (2002). The panels
represent from top to bottom: mass/luminosity versus warp amplitude, total mass in 
solar masses versus warp amplitude, $R_{25}$ versus warp amplitude, absolute 
magnitude versus warp amplitude, Mass/luminosity versus warp asymmetry, total 
mass in solar masses versus warp asymmetry, R$_{25}$ versus warp asymmetry 
for optical sample, absolute magnitude versus warp asymmetry and amplitude 
versus warp asymmetry. The points represent each galaxy in the sample 
and the line is the average of the galaxies with amplitude $>3$,
taking a given width of the bin in the $x$ axis.
}
\label{Fig:figopt}
\end{figure*}

\begin{figure*}
\begin{center}
\mbox{\epsfig{file=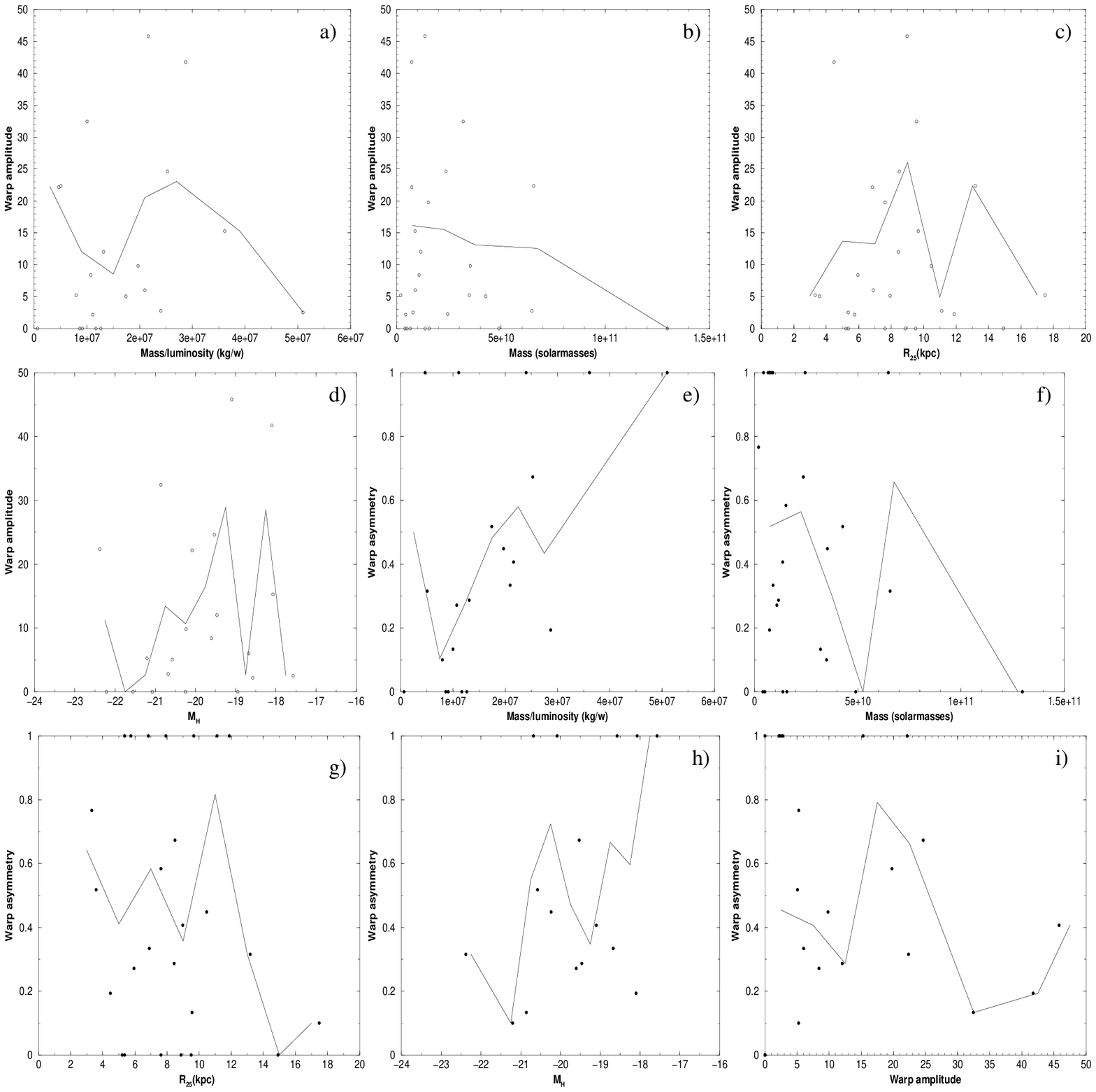,width=16.cm}}
\end{center}
\caption{Radio data from Garc\'\i a-Ruiz (2001). The panels
represent from top to bottom: mass/luminosity versus warp amplitude, total mass in 
solar masses versus warp amplitude, $R_{25}$ versus warp amplitude, absolute 
magnitude versus warp amplitude, mass/luminosity versus warp asymmetry, total 
mass in solar masses versus warp asymmetry, R$_{25}$ versus warp asymmetry 
for optical sample, absolute magnitude versus warp asymmetry and amplitude 
versus warp asymmetry. 
The points represent each galaxy in the sample 
and the line is the average of the galaxies,
taking a given width of the bin in the $x$ axis.}
\label{Fig:figrad}
\end{figure*}

\section{Analysis of the correlations}
\label{.results}

With the information available in the tables we can determine
$R_{25}$ (kpc) from the angular size and the distance,
and the mass $M={R_{25}v^2}/{G}$. The luminosity (or absolute
magnitude) is also immediately derived once we know the apparent magnitude and
the distance. We define the amplitude as the $\frac{1}{2}(EastWA+WestWA)$
and the asymmetry as $|EastWA-WestWA|/(EastWA+WestWA)$.
In this section, we analyse the correlations among the different
quantities.

Our results are represented in Figs \ref{Fig:figopt} and
$\ref{Fig:figrad}$ for optical and radio warps respectively. 
For each one, we have two different sets
of plots, graphs of warp amplitudes and graphs of the warp asymmetries 
against intrinsic parameters of the galaxies.
The following parameters are represented:
\begin{itemize}
\item Warp amplitude against the mass--luminosity relation. The total mass in kilograms that was calculated with the
maximum of the
rotational velocity curve, the radius of the isophote with 25 mag/arcsec$^{2}$
in kpc and the absolute magnitude in $H$ for the reasons given in \S
\ref{.data}. 
\item Warp asymmetry against the same quantities.
\item The relation between the warp amplitude and the asymmetry of the warps
comparing their east and west wings.
\end{itemize}

In Figs $\ref{Fig:figopt}$ and $\ref{Fig:figrad}$ are represented the relations between the parameters of the warp (amplitude and
asymmetry) against: mass-luminosity relation (Fig. $\ref{Fig:figopt}$a and $\ref{Fig:figopt}$e for its relation with the warp amplitude and the warp
asymmetry respectively; and $\ref{Fig:figrad}$a and $\ref{Fig:figrad}$e), the total mass (Fig. $\ref{Fig:figopt}$b and $\ref{Fig:figopt}$f; and $\ref{Fig:figrad}$b and $\ref{Fig:figrad}$f),
$R_{25}$ 
(Fig. $\ref{Fig:figopt}$c and $\ref{Fig:figopt}$g; and $\ref{Fig:figrad}$c and g$\ref{Fig:figrad}$), the absolute magnitude in $H$ (Fig. $\ref{Fig:figopt}$d and $\ref{Fig:figopt}$h; and
$\ref{Fig:figrad}$d and h$\ref{Fig:figrad}$). Finally, Figs $\ref{Fig:figopt}$i and $\ref{Fig:figrad}$i represent the relation between the warp amplitude and the
warp asymmetry. 
The points represent each galaxy of the sample of S\'anchez-Saavedra et al.
(1990, 2002) and Garc\'\i a-Ruiz (2001). 
The number of points in each plot depends on the number of available galaxies
with information on the two variables represented.
The solid line represents the average of the warp amplitude and warp asymmetry 
respectively along the $x$ axis and was determined with data for the
warp amplitude between 3 and 30 for optical data. We try to avoid galaxies with a small warp that might introduce errors in the measurements. In the case of
the radio data all the measurements are presented in the average representation. Here, we have calculated the average value (solid line) with warp amplitudes 
between 0 and 50 for two reasons: there are fewer
galaxies at this wavelength and we cannot discard any of them; it is easier to measure the warp in the radio than in the optical bands. In the case of the warp asymmetry graphs, the
average value is between 0 and 1. In all cases, ten points along the $x$ axis have been used to fit
this value.
The total number of galaxies is
228 for the optical data and 26 for the radio data, but sometimes there is no parameter (in $H$ band because 2MASS is not yet
complete, the rotational velocity, etc.) for all galaxies (see Tables 1 and
2). In these cases, the figures show a lower number of galaxies. 
The data displayed in the figures show the following behaviour:

\begin{itemize}

\item
In the relation between mean warp angle and mass--luminosity relation, we have not 
found any correlation in either  the visible or the radio. Points are 
distributed in all ranges of M/L. Large oscillations in the average value 
(solid line) reinforce this conclusion. In figure 2a there is a fall but is due
to only one point with $M/L\sim
5\times 10^{7}$.

\item However, in the representations of the amplitude versus total
mass, $R_{25}$ (both parameters are related since larger galaxies are more massive) 
there is a slight anticorrelation, more pronounced in the visible and more conspicuous in the plot of amplitude vs. mass. 
Higher values of mass and $R_{25}$
have on average a smaller warp angle. There is an absence of galaxies in all the figures
for high-mass/radius galaxies and high-amplitude warps. For instance, we can see that all
the galaxies with mass greater than 4--5$\times$ 10$^{11}\ M_\odot$ tend to have smaller warps.
This anticorrelation is clearer in Fig. $\ref{Fig:figopt}$b, where the
solid line falls for
large masses. This behaviour is not so clear in Figure $\ref{Fig:figopt}$c because we have a large
concentration of points for small $x$ values and tends to smooth the
fall of the solid line. There are small values of 
$x$ and a low number of galaxies in the radio data (Figures 2b, 2c), so no correlation can be seen in this 
region. In general, galaxies from the Garc\'\i a-Ruiz (2001) sample are nearer and smaller.

\item In asymmetry representations, we see almost the opposite behaviour. 
There are no apparent correlations with mass or radius 
(see Figures $\ref{Fig:figopt}$f and $\ref{Fig:figopt}$g).
There are many oscillations in the average value for Figures 1f, 1g, 2f and 2g, and there is no clear tendency for either 
the mass or the radio to grow in either sample. But there
is a clear anticorrelation in the mass--luminosity relation and in
the absolute magnitude representation for visible warps. For
Figures 1e and 1h there is a slight drop around 10$^{7}$ kg W$^{-1}$ and $-$21 
mag, respectively. For larger M/L ratios, the galaxies have less
asymmetry in their warps (see Figure $\ref{Fig:figopt}$e). For low M/L values, 
the galaxies have a high dispersion 
in asymmetry of between 0 and 0.4. The solid line represents this behaviour. 
In Figures $\ref{Fig:figrad}$e and $\ref{Fig:figrad}$h we cannot see this 
anticorrelation, or, if there is one, it would be opposite to the trend in the visible warps. 
If we consider the last points with warp
asymmetry equal to 1 and a large M/L ratio (see Fig. $\ref{Fig:figrad}$e), the solid
line tends to rise. There are not many
points in this region and the total sample is very poor. The optical data are more complete. The same
behaviour is seen in  Figure $\ref{Fig:figrad}$h.

\item In Figures $\ref{Fig:figopt}$i and $\ref{Fig:figrad}$i,
the asymmetry of the warp amplitude in the east and west side 
against warp amplitude has been represented. For larger 
warp amplitudes we have lower asymmetry. An analogous result is shown in
Garc\'\i a-Ruiz (2001), who used only radio data and found that warp
amplitudes are more prominent than in optical images.
We must bear in mind that the
errors in the measurements are proportional to $\sqrt{2}S_a/A$, 
where $S_a$ is the error in the measurements
and $A$ is the average value. 
This means that for lower amplitudes the error will be larger, and this
could introduce  scatter in the results. In any case, the average
value of the asymmetry should not be affected by this scatter, 
so we can tentatively talk about the 
detection of an anticorrelation between both variables.

\end{itemize}

All these relations are subject to the authenticity of the warp
characteristics measured by S\'anchez-Saavedra et al. (1990, 2002)
and Garc\'\i a-Ruiz (2001), especially for the visible warps, since
these are more likely to be confused with other features
(spiral arms, for instance). Nonetheless, the possible contamination,
if reasonably small (no more than 20\% of the sample), 
would only introduce some noise in the correlations.
Unless most of the data are wrong, it cannot be expected that
the present features are caused by this contamination.

\section{Discussion and conclusions}
\label{.discusion}

In our analysis of the correlations between warp characteristics and
other parameters of the galaxies we find some trends of correlation
or anticorrelation in some cases or nothing in other cases.
The number of galaxies is not very large, so possible minor
systematic errors in the parameters are not totally discarded, 
and the dispersion of values is large, so the correlations among the different 
parameters is not perfect (we have no correlation factor close to 1). 
In any case, we think that these relations reveal some real characteristics
which  can be tentatively examined as follows:

\begin{itemize}
\item There is a slight anticorrelation between the amplitude of
the warp in both directions (average between the east and west amplitudes) and
parameters such as the total mass, luminosity and radius.
We believe that this was to be expected for any mechanism that
produces a warp as a reaction to an external torque, whatever it
 its origin (gravitational torque, magnetic torque, accretion torque;
see L\'opez-Corredoira et al. 2002).
More massive  (i.e. generally larger and more
luminous) galaxies have a more massive disc which forces the warped
rings of the disc to collapse towards the flat disc. 
The more massive  the disc is,
the larger are the internal countertorques of the disc and  
 smaller the amplitude (L\'opez-Corredoira et al. 2002).

\item There is no correlation between the amplitude of
the warp and the mass--luminosity relation. This negative result is indeed
very informative. If the halo were the predominant effect in the
dynamics responsible of the formation of the warp, we would expect
a larger amplitude for higher mass/luminosity ratios (a larger fraction
of dark matter embedded in the halo). Either the rotation curve velocity
is not related to the total mass of the galaxy or the warp amplitude
is  independent of the relative proportion of halo mass. 
Bosma (1991) found that galaxies with small dark halo core 
radii (as determined from  rotation curve decomposition) are less 
likely to be warped, but this could be due to an indirect dependence
on the scales. The fact here is that larger fraction of dark mass in
the galaxy do not relate to the amplitude of the warps.

\item There is no correlation between the asymmetry of the warp (differences
between the east and west amplitudes)
and the total mass and radius.
This means that the reasons for the asymmetry are mainly external
(satellites, accretion, intergalactic magnetic fields) and are independent
of the size of the galaxy.

\item There is an anticorrelation between the asymmetry of the warp and the
mass--luminosity relation and perhaps also with the luminosity of the 
stellar population in the disc, but only for optical warps in both cases. 
In the radio there is no clear correlation, or, if there is one, 
it is opposite to the behaviour in the
visible. This result is somewhat puzzling.
It seems to indicate that the halo is responsible in some degree
for the symmetry of the warp in the inner part, which is visible in 
the optical.
The luminosity density or mass density in the disc would also be related with 
the degree of asymmetry
(since there is no correlation with the radius and there is an anticorrelation
with the total luminosity, it seems that the luminosity density is the
factor to be related with the symmetry).
The more massive the halo is with respect the rest of the galaxy and
the more luminous  the disc is (within a constant radius), 
the more symmetric it is in its inner parts. However, in its outer
parts, visible in the radio, the symmetry seems to be independent of these
factors. It seems that the forces
which produce the asymmetry in the warp (interaction with other
satellites, combination of U- and S-warps, etc.) are predominant
in the most external radii with respect the halo forces, which tend
to produce the symmetry; that is, the asymmetry due to external forces
is effected at larger radii for
smaller dark mass fractions. As a matter of fact, we have a clear example of
this behaviour in our own Galaxy: the gas warp observed in the radio
(Burton 1988) is clearly symmetric for $R<1.6\ R_\odot $ but 
asymmetric for $R>1.6\ R_\odot $
($R_\odot\approx 8$ kpc; L\'opez-Corredoira et al. 2000). In our Galaxy, this radius of transition is   equal
to 13 kpc, which is precisely the value of $R_{25}$ (Goodwin et al. 1998).
A tentative explanation for this would be that the
halo mass distribution, reflected in the rotation curves,
plays a major role for  $R<R_{25}$; outer
rotation curves are not caused by the presence of a massive halo
but have another explanation (magnetic fields, MOND, etc.;
see Battaner \& Florido 2000). In any case, if a very massive
halo existed well beyond $R_{25}$ it is clear that  the asymmetry 
could not be reduced as for $R<R_{25}$ so its dynamical effects
must be negligible with respect to the forces that produce the asymmetries.

\item Galaxies with larger amplitudes are 
more symmetric. This is another observational fact that must be
accounted for by any theory which tries to explain  asymmetric warps.
If we interpret the asymmetries as a superposition of S- and
U-warps, we regard the relative contribution of the U-warp
as lower for larger absolute values of S-warps. For instance, 
in the theory of the accretion of intergalactic matter on to the disc
(L\'opez-Corredoira et al. 2002) this would mean that large warps were 
 produced only when the direction of the infalling flow is
far from the galactic pole, and this provides also small asymmetries.
If the asymmetries were produced by the presence of a companion
galaxy, that would mean that the typical gravitational forces
are comparable to the forces that produce S-warps with small amplitudes,
and that they become unimportant for large S-warps.

\end{itemize}

Summing up, we think that the  correlations analysed here can give
us some clues about the predominant mechanism for the formation of
warps in spiral galaxies. At present, the data seem to indicate
that the role of the halo is  important only in making 
S-warps at $R<R_{25}$ more symmetric, and 
that  asymmetries are more important
in less warped galaxies. This favours scenarios in which
the halo is not very important in the formation of S-warps, especially
radio S-warps and is in agreement with theories that
identify the origin of the warps as directly related to external
 (intergalactic) factors without the mediation of the halo.
The origin of the asymmetries in the warps might be different
from the mechanism of S-warps, and in such a case the halo could
play a role, only in the inner region.

In the introduction, we have described four different theories to explain the formation of warps. The present
results cannot give a definitive answer about which is the correct one. Our goal in the present paper is just to
present observational results, not to defend or deny a particular theory. In any case, as a consequence of these results,
a few words can be added for the comparisons between theories and observations: 
\begin{itemize}
\item Gravitational interaction with a satellite: can be the mechanism, but only if 
there are nearby satellites massive enough to produce the observed warps without any amplification of the halo as an
intermediary, which seems not to be the case in many warped galaxies (for instance, Milky Way and
many apparently isolated galaxies).
\item Intergalactic magnetic field: is in general consistent with the
present results. A slight anticorrelation between asymmetry and
mass/luminous relation is appreciated in Fig. \ref{Fig:figopt}e in $R<R_{25}$. 
Under this hypothesis, the asymmetry could be due to inhomogeneous 
intergalactic fields or to other perturbative effects, even of 
non-magnetic nature. If the asymmetries are really driven by a magnetic 
field mechanism, the found anticorrelation would suggest that this mechanism 
would also have an influence on rotation curves.
\item Misalignment of the halo: should produce some correlations of the warp amplitude with the mass-luminosity ratio,
which are not observed. Hence, unless an explanation can be found for this no-correlation, it seems that this theory should be
discarded.
\item Direct accretion of intergalactic medium onto the disc: explains the present results, but need either 
to have a dark matter halo within $R<R_{25}$ which has some effect on the amplitude of U-warps, or
flat rotation curves are produced by the same matter accretion onto the disc which is responsible of the U-warps.
\end{itemize}

These are just temptative interpretations in the light of the present results. It is also possible
that several mechanism can be present in the warp formation at the same time.
With further data for more galaxies, at higher resolution, and with a more detailed theoretical analysis
of the different hypotheses to fit the observations, 
these results and interpretations can be corroborated and/or improved. 
New work with optical, infrared and radio data could be 
very useful for confirming the present trend and to 
reduce the dispersion of values.

{\it Acknowledgments:}
Thanks are given to Victor P. Debattista and A. Guijarro. 
This article makes use of data products from 2MASS, which is
a joint project of the Univ. of Massachusetts and the Infrared Processing
and Analysis Center, funded by the NASA and the NSF. This work has been supported by
"Cajacanarias" (Canary Islands, Spain) and the project AYA2000-2046-Co2-02 of the 
spanish MCYT.

\appendix

\begin{table*}
\begin{center}
\begin{tabular}{llllllll}\hline

PGC/NGC  &   East WA & West WA&  $cz$ & $V_{\rm rot}$ & log $D_{25}$ & $m_{H}$ &  $d$  \\ 
& &  & (km s$^{-1}$) & (km s$^{-1}$)& 0.1 arcmin &  &  (Mpc)     \\  \hline
\hline
PGC474  &   13 &  12  &  1542 $_{i)}$  &139 $_{i)}$ &  1.53  & ----  &    20.6			\\ 
PGC627  &   13 &  13  &  1495 $_{a)}$  & 77 $_{1)}$ &  1.39  & ----  &    20.0$^*$		\\ 
PGC725  &    19 &  17  &  6004 $_{a)}$  &226 $_{1)}$ &  1.34  & ---- &    80.0   		\\ 
PGC1851  &    20 &  14  &  1596 $_{a)}$  &233 $_{1)}$ &  1.92  &  7.6 &    21.3  		\\ 
PGC1942  &    8  &  13  &  7110 $_{h)}$  &108 $_{h)}$ &  1.41  & 10.1 &    94.8			\\ 
PGC1952  &    15 &  12  &  2626 $_{a)}$  &205 $_{1)}$ &  1.45  & ---- &    35.0 		\\  
PGC2228  &    13 &  12  &  3043 $_{h)}$  &115 $_{1)}$ &  1.31  & 11.6 &    40.6  		\\ 
PGC2482  &    15 &  17  &  3946 $_{h)}$  &291 $_{h)}$ &  1.45  &  9.2  &    52.6  		\\  
PGC2789  &   12 &  15  &  241  $_{g)}$  &204 $_{4)}$ &  2.43  &  4.5  &    3.4  $^*$ 		\\ 
PGC2800  &    13 &  13  &  5765 $_{h)}$  &219 $_{h)}$ &  1.24  & ---- &    76.9			\\ 
PGC2805  &    25 &  17  &  1345 $_{c)}$  & 59 $_{s)}$ &  1.48  & ---- &    17.9			\\ 
PGC3743  &   14 &  12  &  2290 $_{a)}$  &172 $_{a)}$ &  1.57  & ----  &    30.5    		\\ 
PGC4440  &    13 &  20  &  3552 $_{a)}$  &198 $_{1)}$ &  1.41  &  9.8 &    47.4   		\\ 
PGC4912  &    17 &  25  &  5883 $_{a)}$  &234 $_{1)}$ &  1.29  & 10.1 &    40.6  		\\ 
PGC5688  &    9  &  5	&  5431 $_{h)}$  &255 $_{h)}$ &  1.34  &  9.8  &    72.4     		\\ 
PGC6966  &   4  &  12  &  5005 $_{h)}$  &257 $_{1)}$ &  1.48  & ----  &    66.7     		\\ 
PGC7306  &    13 &  13  &  4443 $_{h)}$  &136 $_{h)}$ &  1.29  & ---- &    59.2      		\\ 
PGC7427  &    6  &  13  &  5530 $_{a)}$  &178 $_{1)}$ &  1.27  & ---- &    73.7    		\\ 
PGC8326  &    7  &  8	&  8133 $_{a)}$  &312 $_{1)}$ &  1.40  & 10.2 &   108.4  		\\ 
PGC8673  &   9  &  10  &  1890 $_{h)}$  & 96 $_{1)}$ &  1.34  & ----  &    25.2 		\\ 
PGC9582  &    5  &  11  &  4773 $_{h)}$  &294 $_{1)}$ &  1.33  & ----  &    63.6    		\\ 
PGC10965  &    7  &  7   &  2065 $_{a)}$  &160 $_{1)}$ &  1.57  & ----  &    27.5		\\ 
PGC11198  &    25 &  20  &  4495 $_{h)}$  &211 $_{h)}$ &  1.45  & ---- &    59.9      		\\ 
PGC11595  &    18 &  18  &  1391 $_{a)}$  & 83 $_{1)}$ &  1.47  & 12.0 &    18.5   		\\ 
PGC11659  &    10 &  10  &  5529 $_{a)}$  &255 $_{1)}$ &  1.36  &  9.8  &    73.7     		\\ 
PGC11851  &   9  &  6	   &  1318 $_{a)}$  &116 $_{1)}$ &  1.53  & 10.2  &    17.5  		\\ 
PGC12521  &    15 &  10  &  3949 $_{a)}$  &178 $_{1)}$ &  1.34  & 10.5 &    52.6  		\\ 
PGC13171  &    14 &  10  &  1812 $_{h)}$  &109 $_{h)}$ &  1.40  & 10.1 &    24.1     		\\ 
PGC13458  &    9  &  6   &  1068 $_{a)}$  &150 $_{1)}$ &  1.59  &  9.2 &    14.2     		\\ 
PGC13569  &    0  &  0   &  1638 $_{a)}$  & 65 $_{1)}$ &  1.36  & 11.2 &    21.8     		\\ 
PGC13646  &    12 &  8   &  2168 $_{c)}$  &168 $_{1)}$ &  1.50  & ----  &    28.9		\\ 
PGC13727  &   15 &  14  &  1179 $_{a)}$  &193 $_{1)}$ &  1.88  &  8.3  &    15.7     		\\ 
PGC13809  &   6  &  0	&  1882 $_{a)}$  &131 $_{1)}$ &  1.69  &  9.4  &    25.0      		\\ 
PGC13912  &   9  &  9	   &  980  $_{a)}$  &120 $_{1)}$ &  1.63  & ----  &    ----		\\ 
PGC14071  &   9  &  9	&  1050 $_{a)}$  & 84 $_{h)}$ &  1.40  & ----  &    14.0		\\ 
PGC14190  &    7  &  16  &  1279 $_{a)}$  & 95 $_{1)}$ &  1.56  & ----  &    17.0		\\ 
PGC14255  &    13 &  13  &  1291 $_{a)}$  & 95 $_{1)}$ &  1.28  & 11.3 &    17.2   		\\ 
PGC14259  &    9  &  13  &  4111 $_{a)}$  &175 $_{1)}$ &  1.43  & 10.8 &    54.8   		\\ 
PGC14337  &    0  &  0   &  5386 $_{a)}$  &182 $_{1)}$ &  1.31  & 10.7 &    71.8    		\\ 
PGC14337  &   0  &  0	   &  5386 $_{a)}$  &182 $_{1)}$ &  1.31  & 10.7  &    72.0  		\\ 
PGC14397  &    8  &  8   &  1094 $_{a)}$  &190 $_{1)}$ &  1.72  & ---- &    14.6		\\ 
PGC14824  &    7  &  10  &  1359 $_{c)}$  & 92 $_{1)}$ &  1.59  & ----  &    18.1		\\ 
PGC15455  &    9  &  8   &  1851 $_{a)}$  &120 $_{1)}$ &  1.44  & ---- &    24.6      		\\ 
PGC15635  &    9  &  17  &  4852 $_{h)}$  &288 $_{h)}$ &  1.52  & ---- &    64.6		\\ 
PGC15654  &    10 &  10  &  4759 $_{h)}$  &225 $_{h)}$ &  1.37  & ---- &    63.4    		\\ 
PGC15674  &    14 &  15  &  3705 $_{h)}$  &202 $_{h)}$ &  1.27  & 10.0 &    49.4      		\\ 
PGC15749  &    20 &  20  &  1678 $_{a)}$  & 92 $_{1)}$ &  1.38  & ----  &    22.4		\\ 
PGC16144  &    0  &  0   &  2819 $_{h)}$  &155 $_{h)}$ &  1.35  & 10.7  &    37.6      		\\ 
\hline
\end {tabular}
\end{center}
{{\bf Table 1}: Optical data. Columns in the table represent: name of the galaxy, warp amplitude in the west and east side of
the galaxy, redshift, maximum rotation velocity, log(D$_{25}$), H magnitude and distance. The references of each
value are: *, Huchtmeier et al. (1989); a), Mathewson et al. (1996); b),Di Nella et al. (1996); c), Da Costa et
al. (1998); d), Saunders et al. (2000); e), Longmore et al. (1982); f), Haynes et al. (1998); g), Huchtmeier
et al. (1985); h), Theureau et al. (1998); i), Fisher et al. (1981); ii), Tully (1988); j), Richter et al. (1987); k), Davies et
al. (1989); l), de Vaucouleurs et al. (1991), ll), Thuan et al. (1981); m), Strauss et al. (1992); 
n), Fairall et al. (1988); o), Staveley-Smith et al. (1987); p), Dressler et al. (1991); q), Fairall et al. 
(1991); r), Chengalur et al.(1993); s), Tifft et al. (1988); t), Giovanelli et al. (1997); u), Fairall et al. 
(1992); v), Bottinelli et al. (1993); w), Loveday et al. (1996); 1), Mathewson et al. (1992); 2), 
Staveley-Smith et al. (1988); 3), Reif et al. (1982); 4), Corradi et al. (1991);
 5), Banks et al. (1999).}
\end {table*}

\begin{table*}
\begin{center}
\begin{tabular}{llllllll}\hline

PGC/NGC  &   East WA & West WA&  $cz$ & $V_{\rm rot}$ & log $D_{25}$ & $m_{H}$ &  d  \\ 
& &  & (km s$^{-1}$) & (km s$^{-1}$)& 0.1 arcmin &  &  (Mpc)     \\  \hline
\hline

PGC16168  &    7  &  7   &  4577 $_{h)}$  &170 $_{h)}$ &  1.22  & ---- &    61.0    		\\ 
PGC16199  &    12 &  12  &  1169 $_{a)}$  & 87 $_{1)}$ &  1.44  & ---- &    15.6    		\\ 
PGC16636  &    0  &  0   &  4329 $_{h)}$  &199 $_{h)}$ &  1.49  & ----  &    57.7     		\\ 
PGC17056  &    10 &  10  &  2828 $_{a)}$  &172 $_{1)}$ &  1.24  & ----  &    37.7		\\ 
PGC17174  &    0  &  0   &  1755 $_{a)}$  &158 $_{h)}$ &  1.51  & ----  &    23.4		\\ 
PGC17969  &   10 &  10  &  2382 $_{h)}$  &145 $_{1)}$ &  1.34  & 10.5 &     31.8   		\\ 
PGC18437  &    6  &  5   &  1228 $_{a)}$  &137 $_{1)}$ &  1.60  &  9.6  &    16.4    		\\ 
PGC18765  &   0  &  0	   &  1696 $_{i)}$  &143 $_{a)}$ &  1.52  &  9.8  &    22.6		\\ 
PGC19996  &   5  &  5	   &  2681 $_{h)}$  &166 $_{f)}$ &  1.35  & 10.1  &    35.7		\\ 
PGC21815  &    6  &  6   &  1131 $_{b)}$  & 98 $_{b)}$ &  1.67  &  8.4  &    15.1		\\ 
PGC21822  &    17 &  7   &  3237 $_{h)}$  &245 $_{h)}$ &  1.48  &  9.3 &    43.1		\\ 
PGC22272  &    0  &  0   &  1558 $_{h)}$  &130 $_{h)}$ &  1.45  & 11.1  &    20.8      		\\ 
PGC22338  &   8  &  9	   &  1119 $_{ii)}$ &148 $_{m)}$ &  1.80  & ----  &    14.9 $^*$     	\\ 
PGC22910  &    13 &  12  &  5954 $_{a)}$  &223 $_{1)}$ &  1.41  & ---- &    79.4    		\\ 
PGC23558  &    20 &  20  &  1776 $_{h)}$  & 91 $_{h)}$ &  1.27  & ---- &    23.7		\\ 
PGC23992  &    15 &  17  &  4533 $_{a)}$  &190 $_{1)}$ &  1.15  & ---- &    60.4    		\\ 
PGC24685  &   6  &  8	   &  4570 $_{a)}$  &308 $_{1)}$ &  1.48  &  9.4  &    60.9	 	\\ 
PGC25886  &    9  &  9      &  1838 $_{h)}$  &256 $_{h)}$ &  1.63  & ---- &    24.5		\\ 
PGC25926  &    3  &  3   &  2178 $_{a)}$  &158 $_{a)}$ &  1.65  & ---- &    29.0    		\\ 
PGC26561  &    10 &  10  &  1640 $_{a)}$  &245 $_{1)}$ &  1.75  & ---- &    21.8    		\\ 
PGC27135  &    0  &  0   &  929  $_{i)}$  &100 $_{1)}$ &  1.86  & ----  &     6.40 $^*$     	\\  
PGC27735  &   13 &  13  &  4449 $_{a)}$  &171 $_{1)}$ &  1.28  & ----  &    59.3     		\\  
PGC28117  &   17 &  27  &  4315 $_{a)}$  &170 $_{1)}$ &  1.41  & ----  &    57.5    		\\  
PGC28246  &    17 &  20  &  2893 $_{a)}$  &183 $_{1)}$ &  1.50  & ---- &    38.5   		\\  
PGC28283  &    0  &  0   &  2868 $_{h)}$  &220 $_{h)}$ &  1.59  & ----  &    38.2		\\  
PGC28778  &    8  &  8   &  2697 $_{a)}$  &154 $_{i)}$ &  1.40  & ----  &    36.0		\\  
PGC28840  &   7  &  8	   &  2802 $_{a)}$  &123 $_{1)}$ &  1.53  & ----  &    37.3  		\\  
PGC28909  &    0  &  0      &  2520 $_{a)}$  &208 $_{1)}$ &  1.83  & ---- &    33.6	 	\\  
PGC29691  &   0  &  0	   &  2840 $_{h)}$  &142 $_{1)}$ &  1.35  & ----  &    37.9		\\  
PGC29716  &    0  &  0   &  2526 $_{v)}$  &161 $_{k)}$ &  1.59  & ----  &    33.7		\\  
PGC29743  &    9  &  9   &  2603 $_{j)}$  &164 $_{1)}$ &  1.50  & ----  &    34.7		\\  
PGC29841  &   0  &  0	   &  3603 $_{h)}$  &185 $_{1)}$ &  1.32  & 10.5 &     48.0	 	\\  
PGC30716  &   0  &  0	   &  3138 $_{a)}$  &160 $_{1)}$ &  1.30  & ----  &    41.8	 	\\  
PGC31154  &    0  &  0      &  3608 $_{a)}$  &254 $_{1)}$ &  1.35  & ---- &    48.1	 	\\  
PGC31426  &    10 &  6   &  5042 $_{c)}$  &290 $_{u)}$ &  1.41  & ---- &    67.2		\\  
PGC31677  &    10 &  0      &  3756 $_{h)}$  &204 $_{1)}$ &  1.50  & ---- &    50.1 		\\  
PGC31723  &    11 &  0   &  4152 $_{h)}$  &169 $_{1)}$ &  1.32  & ---- &    55.4    		\\  
PGC31919  &    0  &  0   &  1032 $_{a)}$  & 60 $_{1)}$ &  1.46  & 11.8  &    13.8    		\\  
PGC31995  &   8  &  8	&  2932 $_{a)}$  &185 $_{1)}$ &  1.45  & ----  &    39.1      		\\  
PGC32271  &    6  &  7      &  3047 $_{h)}$  &218 $_{h)}$ &  1.54  & ---- &    40.6		\\  
PGC32328  &    0  &  0      &  5704 $_{a)}$  &245 $_{1)}$ &  1.30  & ---- &    76.0	 	\\  
PGC32550  &   6  &  0	   &  3108 $_{j)}$  &114 $_{a)}$ &  1.54  & ----  &    41.4		\\  
PGC35861  &    25 &  25  &  2702 $_{h)}$  &241 $_{h)}$ &  1.53  & ----  &    36.0   		\\  
PGC36315  &    10 &  6   &  3701 $_{a)}$  &136 $_{1)}$ &  1.21  & ---- &    49.3    		\\  
PGC37178  &    0  &  0   &  2013 $_{a)}$  &141 $_{1)}$ &  1.60  &  9.7 &    26.8     		\\  
PGC37243  &    5  &  4   &  2944 $_{a)}$  &176 $_{1)}$ &  1.42  & ----  &    39.2    		\\  
PGC37271  &    7  &  6   &  1702 $_{a)}$  &123 $_{a)}$ &  1.64  & ---- &    22.7    		\\  
PGC37304  &   9  &  9	&  5715 $_{a)}$  &254 $_{1)}$ &  1.36  & ----  &    76.2      		\\  
PGC37334  &   0  &  0	   &  2889 $_{b)}$  &162 $_{b)}$ &  1.42  & ----  &    38.5  		\\  
PGC38426  &    12 &  11  &  4476 $_{c)}$  &198 $_{1)}$ &  1.31  & ----  &    59.7   		\\  
PGC38464  &   5  &  9	   &  1728 $_{a)}$  &121 $_{1)}$ &  1.38  & ----  &    23.0		\\  
PGC38841  &    0  &  0   &  3133 $_{a)}$  &150 $_{h)}$ &  1.31  & ----  &    41.8		\\  
PGC40023  &    0  &  0   &  2940 $_{a)}$  &237 $_{j)}$ &  1.50  & ----  &    39.2		\\  
PGC40284  &   17 &  19  &  2002 $_{a)}$  &176 $_{1)}$ &  1.49  &  9.3  &    26.7    		\\  
PGC42684  &    0  &  0   &  5502 $_{a)}$  &221 $_{1)}$ &  1.32  & ---- &    73.4     		\\  
PGC42747  &   13 &  20  &  3210 $_{a)}$  &145 $_{1)}$ &  1.42  & 11.2 &     42.8      		\\  
PGC43021  &   0  &  0	   &  5260 $_{a)}$  &278 $_{1)}$ &  1.41  & 10.0 &     70.1	 	\\  
PGC43224  &    10 &  10  &  3211 $_{h)}$  &162 $_{h)}$ &  1.25  & 10.3 &    42.8    		\\  
PGC43313  &    0  &  0   &  3693 $_{c)}$  &209 $_{u)}$ &  1.40  &  10. &    49.2    		\\  
PGC43330  &   14 &  16  &  1408 $_{c)}$  & 60 $_{2)}$ &  1.47  & 10.3  &    18.8  		\\

\hline
\end {tabular}
\end{center}
{Data. Continuation.}
\end {table*}

\begin{table*}
\begin{center}
\begin{tabular}{llllllll}\hline

PGC/NGC  &   East WA & West WA&  $cz$ & V$_{\rm rot}$ & logD$_{25}$ & m$_{H}$ &  d  \\ 
& &  & (Km/s) & (Km/s)& 0.1 arcmin &  &  (Mpc)     \\  \hline

PGC43342  &    0  &  0      &  4459 $_{h)}$  &254 $_{h)}$ &  1.39  & ---- &    59.4		\\  
PGC43679  &    0  &  0   &  2258 $_{i)}$  &106 $_{j)}$ &  1.39  & ---- &    30.1     		\\  
PGC44254  &    0  &  0      &  2839 $_{c)}$  &142 $_{u)}$ &  1.28  & 10.7 &    37.8		\\  
PGC44271  &    0  &  0   &  3376 $_{a)}$  &172 $_{1)}$ &  1.43  & ----  &    45.0		\\  
PGC44358  &    0  &  0   &  1487 $_{c)}$  &114 $_{i)}$ &  1.51  & ---- &    19.8     		\\  
PGC44409  &    0  &  0      &  2173 $_{a)}$  &184 $_{1)}$ &  1.67  & ---- &    29.0 		\\  
PGC44931  &   8  &  7	   &  3812 $_{c)}$  &201 $_{1)}$ &  1.45  & ----  &    50.8	 	\\  
PGC44966  &    0  &  0      &  4995 $_{a)}$  &231 $_{1)}$ &  1.19  & ---- &    66.6	 	\\  
PGC45006  &   9  &  13  &  4527 $_{c)}$  &206 $_{1)}$ &  1.42  & ----  &    60.4     		\\  
PGC45098  &    12 &  9      &  2896 $_{a)}$  &169 $_{1)}$ &  1.46  & ---- &    38.6		\\  
PGC45127  &    10 &  10  &  4007 $_{h)}$  &180 $_{1)}$ &  1.27  & 10.7  &    53.4		\\  
PGC45279  &   14 &  14  &  560  $_{a)}$  &180 $_{a)}$ &  2.31  &  7.5  &    6.7 $^*$	    	\\  
PGC45487  &    0  &  0   &  2621 $_{a)}$  &114 $_{1)}$ &  1.48  & ---- &    34.9     		\\  
PGC45911  &   0  &  0	   &  2754 $_{a)}$  &143 $_{1)}$ &  1.47  & ----  &    36.7		\\  
PGC45952  &   0  &  0	   &  3006 $_{a)}$  &170 $_{1)}$ &  1.38  & ----  &    40.1  		\\  
PGC46441  &   10 &  10  &  2744 $_{d)}$  &191 $_{2)}$ &  1.54  & ----  &    36.6     		\\  
PGC46650  &   4  &  15  &  2566 $_{e)}$  &131 $_{a)}$ &  1.46  & ----  &    34.2  		\\  
PGC46768  &   0  &  0	   &  2256 $_{a)}$  &112 $_{1)}$ &  1.25  & ----  &    30.1	 	\\  
PGC47345  &   7  &  14  &  3604 $_{h)}$  &207 $_{h)}$ &  1.52  & ----  &    48.0		\\  
PGC47394  &   0  &  0	&  1503 $_{a)}$  &251 $_{a)}$ &  1.91  &  8.6  &    20.0  		\\  
PGC47948  &    8  &  9   &  2577 $_{a)}$  &158 $_{1)}$ &  1.40  & ----  &    34.4		\\  
PGC48359  &    0  &  0   &  3631 $_{v)}$  &229 $_{1)}$ &  1.31  & ----  &    48.4		\\  
PGC49129  &    9  &  15  &  141  $_{a)}$  & 47 $_{a)}$ &  1.41  & ---- &    ---- 		\\  
PGC49190  &    17 &  15  &  3931 $_{h)}$  & 93 $_{h)}$ &  1.23  & ---- &    52.4		\\  
PGC49586  &   8  &  8	   &  2760 $_{a)}$  &196 $_{b)}$ &  1.45  & ----  &    36.8	    	\\  
PGC49676  &   11 &  13  &  2663 $_{a)}$  &241 $_{a)}$ &  1.76  &  8.7  &    35.5      		\\  
PGC49836  &    4  &  10  &  2907 $_{a)}$  &153 $_{1)}$ &  1.37  & ---- &    38.8      		\\  
PGC50676  &    14 &  15  &  1541 $_{a)}$  &112 $_{1)}$ &  1.64  & 10.9 &    20.5  		\\  
PGC50798  &    0  &  0      &  3017 $_{a)}$  &164 $_{1)}$ &  1.41  & ---- &    40.2		\\  
PGC51613  &   0  &  0	   &  2245 $_{a)}$  &123 $_{1)}$ &  1.48  & ----  &    29.9	 	\\  
PGC52410  &    0  &  0   &  2869 $_{a)}$  &174 $_{1)}$ &  1.35  & ----  &    38.2    		\\  
PGC52411  &   9  &  9	   &  3420 $_{a)}$  &215 $_{1)}$ &  1.45  &  9.5  &    45.6	 	\\  
PGC52991  &   0  &  0	   &  2945 $_{a)}$  & 99 $_{1)}$ &  1.30  & 11.3 &     39.3	 	\\  
PGC53361  &   0  &  0	   &  4510 $_{a)}$  &152 $_{1)}$ &  1.36  & ----  &    60.1		\\  
PGC54392  &    0  &  0   &  522  $_{m)}$  & 79 $_{5)}$ &  2.05  & ---- &     7.0$^*$	     	\\  
PGC54637  &    9  &  12  &  4655 $_{a)}$  &212 $_{1)}$ &  1.40  & ---- &    62.0    		\\  
PGC56077  &    0  &  0      &  2692 $_{a)}$  &115 $_{1)}$ &  1.30  & ---- &    35.9		\\  
PGC57582  &    0  &  0   &  2044 $_{f)}$  &169 $_{i)}$ &  1.78  & ----  &    27.2      		\\  
PGC57876  &    0  &  0   &  3410 $_{a)}$  &222 $_{1)}$ &  1.59  & ----  &    45.5    		\\  
PGC59635  &    0  &  0   &  1508 $_{a)}$  &101 $_{1)}$ &  1.57  & ---- &    20.1     		\\  
PGC60216  &   15 &  15  &  2859 $_{a)}$  &109 $_{1)}$ &  1.30  & ----  &    38.1   		\\  
PGC60595  &   0  &  0	&  4698 $_{a)}$  &190 $_{1)}$ &  1.39  & ----  &    62.6      		\\  
PGC62024  &    0  &  0      &  3183 $_{a)}$  &201 $_{1)}$ &  1.22  & ---- &    42.4	 	\\  
PGC62706  &    0  &  0   &  3182 $_{a)}$  &133 $_{1)}$ &  1.57  & ---- &    42.4     		\\  
PGC62782  &    8  &  4   &  1841 $_{a)}$  & 83 $_{1)}$ &  1.47  & ---- &    24.5		\\  
PGC62816  &    10 &  10  &  5024 $_{a)}$  &233 $_{1)}$ &  1.25  & ---- &    66.9    		\\  
PGC62922  &    7  &  0   &  4404 $_{a)}$  &280 $_{1)}$ &  1.57  & ----  &    58.7		\\  
PGC62964  &   13 &  13  &  2847 $_{a)}$  &241 $_{4)}$ &  1.62  & ----  &    38.0      		\\  
PGC63395  &    3  &  6   &  1928 $_{a)}$  &117 $_{a)}$ &  1.51  & ---- &    25.7		\\  
PGC63577  &   12 &  12  &  4231 $_{a)}$  &138 $_{1)}$ &  1.29  & ----  &    56.4    		\\  
PGC64597  &   5  &  5	   &  4196 $_{a)}$  &120 $_{a)}$ &  1.34  & ----  &    55.9		\\  
PGC65794  &    9  &  3   &  9150 $_{a)}$  &323 $_{1)}$ &  1.42  & ---- &    122.0   		\\  
PGC65915  &    11 &  8      &  3122 $_{a)}$  &177 $_{1)}$ &  1.53  & ---- &    41.6 		\\  
PGC66530  &    14 &  7      &  3144 $_{a)}$  &266 $_{1)}$ &  1.50  & ---- &    41.9		\\  
PGC66617  &   0  &  0	   &  2715 $_{r)}$  &101 $_{1)}$ &  1.25  & ----  &    36.2	 	\\  
PGC66836  &   0  &  0	   &  797  $_{d)}$  & 73 $_{1)}$ &  1.52  & ----  &    16.2 $^*$     	\\  
PGC67045  &   0  &  0	   &  857  $_{d)}$  & 96 $_{3)}$ &  1.89  &  9.3  &    16.0 $^*$    	\\  
PGC67078  &    0  &  0   &  2479 $_{a)}$  & 85 $_{1)}$ &  1.30  & ----  &    33.0   		\\  
PGC67158  &    0  &  7   &  3400 $_{a)}$  &175 $_{1)}$ &  1.44  & 10.3  &    45.3   		\\  
PGC67904  &   6  &  5	   &  2635 $_{a)}$  &264 $_{1)}$ &  1.78  &  8.5  &    35.1		\\  

\hline
\end {tabular}
\end{center}
{Data. Continuation.}
 \end {table*}

\begin{table*}
\begin{center}
\begin{tabular}{llllllll}\hline

PGC/NGC  &   East WA & West WA&  $cz$ & V$_{\rm rot}$ & logD$_{25}$ & m$_{H}$ &  d  \\ 
& &  & (Km/s) & (Km/s)& 0.1 arcmin &  &  (Mpc)     \\  \hline
\hline

PGC68223  &    11 &  11  &  2847 $_{r)}$  &169 $_{r)}$ &  1.38  & 10.1 &    38.0    		\\  
PGC68389  &    6  &  5   &  1746 $_{a)}$  &174 $_{1)}$ &  1.64  & ---- &    23.3     		\\  
PGC69161  &    19 &  18  &  2091 $_{k)}$  &117 $_{1)}$ &  1.55  & 10.9 &    27.9  		\\  
PGC69539  &    9  &  8   &  1240 $_{a)}$  &102 $_{1)}$ &  1.60  & 10.7  &    16.5   		\\  
PGC69661  &    16 &  11  &  2360 $_{a)}$  &175 $_{1)}$ &  1.48  &  9.8 &    31.5    		\\  
PGC69707  &    0  &  0   &  2364 $_{a)}$  &100 $_{1)}$ &  1.59  & ---- &    31.5		\\  
PGC69967  &    0  &  0      &  3001 $_{a)}$  &148 $_{1)}$ &  1.41  & 10.5 &    40.0 		\\  
PGC70025  &    7  &  7      &  2857 $_{n)}$  &167 $_{f)}$ &  1.50  & ---- &    38.1		\\  
PGC70070  &    0  &  0      &  1681 $_{a)}$  &109 $_{1)}$ &  1.58  & ---- &    22.4		\\  
PGC70081  &    9  &  9   &  1940 $_{a)}$  &240 $_{1)}$ &  1.49  & ----  &    25.9		\\  
PGC70084  &   13 &  7	   &  5041 $_{a)}$  &308 $_{a)}$ &  1.31  & ----  &    67.2		\\  
PGC70324  &    0  &  0   &  1059 $_{a)}$  & 85 $_{a)}$ &  1.62  & 10.1 &    14.1		\\  
PGC71800  &    7  &  0   &  2008 $_{a)}$  &101 $_{a)}$ &  1.24  & ---- &    26.7		\\  
PGC71948  &    0  &  0      &  2876 $_{a)}$  &253 $_{1)}$ &  1.74  & ---- &    38.3	 	\\  
PGC72178  &    4  &  8   &  1489 $_{a)}$  &110 $_{1)}$ &  1.43  & ----  &    19.8		\\  
NGC4013  &   5  &  5     &  834  $_{f)}$  &193 $_{1)}$ &  1.72  &  8.7  &    12.0$^*$      	\\  
NGC1560   &    5  &  5   &  -36  $_{d)}$  & 76 $_{1)}$ &  1.99  &  9.4 &     3.0 $^*$      	\\  
NGC2654   &    8  &  8   &  1347 $_{f)}$  &197 $_{1)}$ &  1.63  & ---- &    22.4      		\\  
NGC2683   &   7  &  7   &  411  $_{f)}$  &275 $_{1)}$ &  1.97  &  6.8  &     5.1 $^*$       	\\  
NGC2820  &   12 &  16  &  3811 $_{o)}$  &210 $_{1)}$ &  1.46  &  9.5  &    50.8    		\\  
NGC2820  &   12 &  16  &  3811 $_{o)}$  &210 $_{1)}$ &  1.46  &  9.5  &    50.8    		\\  
NGC3510   &    10 &  10  &  705  $_{ll)}$ & 83 $_{1)}$ &  1.58  & 11.2 &     9.0 $^*$      	\\  
NGC3628   &   16 &  16  &  843  $_{s)}$  &223 $_{1)}$ &  2.17  &  6.9  &     6.7 $^*$		\\  
NGC4010   &    6  &  6   &  907  $_{i)}$  &118 $_{1)}$ &  1.62  & 10.2 &    11.0 $^*$      	\\  
NGC4565   &   2  &  2   &  1282 $_{t)}$  &259 $_{1)}$ &  2.21  &  6.7  &    10.0		\\  
NGC6045   &   11 &  11  &  9986 $_{a)}$  &258 $_{1)}$ &  1.12  & 10.9  &    133.1   		\\  
NGC6161  &   14 &  12  &  5904 $_{a)}$  &256 $_{1)}$ &  1.29  & 10.2 &     78.7   		\\  
NGC6242  &   13 &  13  &  4620 $_{a)}$  &172 $_{1)}$ &  1.28  & ----  &    61.6    		\\  
NGC7640   &   7  &  7     &  369  $_{f)}$  &110 $_{1)}$ &  2.03  &  9.3  &    9.2 $^*$      	\\  
	
\hline
\hline
\end {tabular}
\end{center}
{Data. Continuation.}
 \end {table*}

\begin{table*}
\begin{center}
\begin{tabular}{llllllll}\hline

UGC&EAST WA& WEST WA& $cz$& V$_{\rm rot}$&logD$_{25}$&m$_{H}$&d \\ 
& &&Km/sg&Km/sg&0.1 arcmin&&Mpc \\ \hline
\hline
1281 &  1.22 &  9.27 &  157 &	50 & 1.65  &----  &   5.1 \\  
2549 &  2.44 &  7.69 & 10355 &  226 &  .83  &12.2  &  36.3 \\ 
3137 &  5.76 &  4.71 &  992 &	93 & 1.55  &*11.4 &  33.8 \\  
3909 & 15.48 &  8.57 &  945 &	77 & 1.37  &12.5  &  24.5 \\  
4278 &  5.06 &  0.00 &  560 &	79 & 1.66  &11.9  &   8.1 \\  
4806 &  0.00 &  5.59 &  1947 &  158 & 1.56  &*10.9 &  21.1 \\ 
5452 & 31.33 &  8.22 &  1342 &  93 & 1.38  &----  &  21.7 \\  
5459 &  5.41 & 14.23 &  1112 &  120 & 1.66  &*10.8 &  15.9 \\ 
5986 & 41.21 &  8.04 &  615 &  109 & 1.84  &10.1  &   8.5 \\  
6126 & 49.85 & 33.65 &  704 &	83 & 1.54  &11.6  &   8.8 \\  
6283 &  6.11 & 10.68 &  719 &	88 & 1.56  &10.7  &  11.3 \\  
6964 & 28.10 & 36.79 &  905 &  120 & 1.59  &10.3  &  16.9 \\  
7089 &  0.00 &  0.00 &  774 &	57 & 1.50  & 8.8  &  11.6 \\  
7090 &  0.00 &  0.00 &  560 &  149 & 1.81  & *8.9 &  10.2 \\  
7125 & 10.33 &  0.00 &  1071 &  59 & 1.64  &----  &  12.6 \\  
7151 &  0.00 &  0.00 &  267 &	64 & 1.78  & 9.9  &   6.0 \\  
7321 &  4.54 &  0.00 &  409 &	94 & 1.74  &----  &  14.9 \\  
7483 &  0.00 &  0.00 &  1248 &  94 & 1.47  &10.9  &  17.6 \\  
7774 & 64.44 & 27.16 &  526 &	80 & 1.48  &12.5  &  20.6 \\  
8246 & 30.57 &  0.00 &  794 &	63 & 1.53  &13.4  &  19.4 \\  
8286 &  8.04 &  4.01 &  407 &	75 & 1.77  &*10.8 &   8.0 \\  
8396 & 44.31 &  0.00 &  945 &	68 & 1.23  &12.1  &  27.5 \\  
8550 &  0.00 &  4.36 &  364 &	57 & 1.48  &12.0  &  13.2 \\  
8709 &  0.00 &  0.00 &  2402 &  194 & 1.71  & 9.3  &  19.8 \\ 
8711 & 15.30 & 29.43 &  1531 &  146 & 1.60  & 9.4  &  22.5 \\ 
9242 &  0.00 &  0.00 &  1436 &  81 & 1.68  &----  &  12.6 \\

 \hline
\end {tabular}
\end{center}
{{\bf Table 2}: Radio data. Columns in the table represent:name of the galaxy, warp amplitude in the west and east side of
the galaxy, redshift, maximum rotation velocity, log(D$_{25}$), H magnitude and distance.The references of each value are: Columns 1, 2, 3 from
Garc\'\i a-Ruiz (2001), the warp amplitudes are in the same units than optical
amplitudes; columns 4, 5 and 6 from Garc\'\i a-Ruiz (2001) and LEDA
database;
column 7 from 2MASS and galaxies with (*) Tormen et al. (1995); column 8 from
Garc\'\i a-Ruiz (2001).}
\end {table*}

\end{document}